# Interference Mitigation and Capacity Enhancement based on Dynamic Frequency Reuse for Femtocell Networks

Md. Tashikur Rahman, Md. Didarul Alam, and Mostafa Zaman Chowdhury
Dept. of Electrical and Electronic Engineering, Khulna University of Engineering & Technology, Bangladesh
E-mail: rahmantashik@gmail.com, didar_2k11@hotmail.com, mzceee@yahoo.com

*Abstract*—Wireless networks employing small cells like femtocells are considered to be the choice of network deployment for 4G or advanced networks. This hierarchical deployment of cells introduces the necessity of effective frequency planning for mitigation of interference between different layers of network. As the scarce spectrum resources are likely to be reused to increase spectral efficiency, interference free signal reception has to be guaranteed to ensure better quality of service (QoS). In this paper we propose a dynamic frequency reuse scheme for the deployment of femtocells within a macrocell with the femtocells reusing the spectrum of neighbouring macrocells. We also provide a protective scheme for cell edge femtocell users as they are vulnerable to interference signals from neighbouring macrocells. A detailed frequency planning is provided to maximize spectral reuse while providing maximum throughput. We compare our proposed scheme with other frequency allocation schemes already described in literature. Simulation results shows that our scheme provides better throughput and ensures lower outage probability.

## I. INTRODUCTION

With added data and multimedia features wireless cellular networks need to provide not only voice services but also a verity of services. This added data traffic is more likely to originate within indoor environment as a study by ABI research have predicted that overall 50% voice and 70% data service request are to be generated from indoor users [1]. Femtocells which are low power, short range base stations are likely to be installed within indoor environment can provide this high requirement of data rate.

With scarce spectrum resource the only viable solution for deployment of femtocells is to effectively reuse the frequency band from macrocell spectrum with proper interference mitigation technique. Authors in [1] have discussed the uplink and downlink scenarios in which both layers of networks can be either the offender or the victim. In this paper we propose a dynamic frequency reuse scheme for femtocell deployment. In our proposed model we considered a macrocell cluster with three macrocells, the femtocells deployed within each macrocell reuse the frequency bands allocated to the neighbouring macrocells (i.e. femtocells within macrocell 1 use the frequency band allocated to macrocell 2 and macrocell 3). To ensure interference free signal reception for cell edge femto user equipment (FUE) we propose formation of a guard region at the cell edge of each macrocell coverage area, in which the femtocells are allocated with frequency bands separated from the macrocell spectrum.

In previous time fractional frequency reuse (FFR) [3] schemes have been proposed, these schemes showed fine interference mitigation capabilities, however they are much dependent on strict partitioning of coverage area which is not optimal under dynamic load variation. Complementary frequency allocation [4] has been proposed which allocates complementary frequencies from the macrocell spectrum to the femtocells. This concept of reverse frequency allocation by changing the uplink and downlink frequencies for macro user equipment (MUEs) and FUEs causes interference between user equipment of different layers. Co-channel, dedicated-channel, and hybrid access [5], [6] modes for femtocell deployment have been proposed, they improve the overall spectrum utilization but interference between different layers of networks result in high outage probability.

The rest of this paper is organized as follows. Section II shows the proposed dynamic frequency reuse scheme with detailed frequency planning. The capacity and outage probability analysis of the proposed scheme are shown In Section III. In Section IV performance evaluation are shown. Finally concluding notes are drawn in Section V.

## II. SYSTEM MODEL

In our proposed model we consider a cluster of three macrocells where three different frequency bands are used. The dynamic frequency reuse (DFR) concept states that the femtocells within coverage area of macrocell 1, reuses the frequency band of macrocell 2 and macrocell 3. However, the FUE at cell-edge are subjected to severe interference from the neighbouring macrocells under dense deployment scenarios. Fig. 1(a) shows the interference caused by the macrocell 2 and macrocell 3 at cell-edge FUEs in macrocell 1. In this context we propose formation of a guard region by allocating dedicated frequency band separated solely for femtocells. To make sure that separating frequency bands from macrocell spectrum does not reduce the total available bandwidth for

MUE, we propose creation of a set of frequency band separated from macrocells from the cluster. Fig. 1(b) shows the frequency bands $B_X$, $B_Y$, and $B_Z$ each separated from the macrocell spectrum. This three frequency bands form a set which is allocated to the femtocells operating in the cell-edge zones of each macrocell. This dedicated frequency bands between two macrocell spectrum answers the problem rising from the adjacent channel interference for the MUEs.

Our proposed model provides detailed frequency division mechanism keeping in mind the number of femtocells deployed in close proximity. Fig. 2(a) shows the sub-band allocation between femtocells under different density of femtocells. For the cell-edge zone femtocells the femto dedicated sub-band which is adjacent to the overlaid macrocell frequency band has the lowest priority i.e. femtocells at cell-edge at macrocell 1 will first use the frequency bands $B_Y$, and $B_Z$ then $B_X$. Parameters used in this section are summarised in Table 1.

TABLE I Basic nomenclature

| Symbol | Definition |
|---|---|
| $B_T$ | The total system-wide spectrum of frequencies |
| $B_{m1}, B_{m2}, B_{m3}$ | Frequency spectrum allocated to macrocells |
| $B_X, B_Y, B_Z$ | Frequency bands for cell-edge femtocells |
| $B_{m4}, B_{m5}, B_{m6}$ | Frequency bands for cell-centre femtocells |
| $B_{f1}, B_{f2}, B_{f3}$ | The total frequency bands allocated to femtocells in macrocell #1, #2, #3, respectively, of a macrocell cluster |
| $B_{fa}, B_{fb}, B_{fc}$ | Actual frequency bands allocated to femtocells |

For the DFR scheme with dedicated frequency band for cell-edge femtocells, the following set of equations can be written;

$$|B_T| = |B_{m1}| + |B_X| + |B_{m2}| + |B_Y| + |B_{m3}| + |B_Z|$$
$$|B_{m4}| = |B_{m5}| = |B_{m6}| = \frac{|B_{m2}| + |B_{m3}|}{3}$$
$$B_{f1} = |B_{m2}| + |B_{m3}| + |B_X| + |B_Y| + |B_Z| \quad (1)$$
$$B_{f2} = |B_{m1}| + |B_{m3}| + |B_X| + |B_Y| + |B_Z|$$
$$B_{f3} = |B_{m1}| + |B_{m2}| + |B_X| + |B_Y| + |B_Z|$$

$$B'_{f1,innerfemto} = \begin{cases} |B_{m4}| + |B_{m5}| + |B_{m6}|, & \text{3 interfering femtocell} \\ |B_{m2}| + |B_{m3}|, & \text{2 or no interfering femtocell} \end{cases} \quad (2)$$

$$B'_{f1,celledgefemto} = \begin{cases} |B_X| + |B_Y| + |B_Z|, & \text{3 interfering femtocell} \\ |B_Y| + \frac{|B_X|}{2} + |B_Z| + \frac{|B_X|}{2}, & \text{2 interfering femtocell} \\ |B_Y| + |B_Z|, & \text{no interfering femtocell} \end{cases} \quad (3)$$

$$B_{m1} \cap B_{f1} = \emptyset, \ B_{m2} \cap B_{f2} = \emptyset, \ B_{m3} \cap B_{f3} = \emptyset$$
$$B_{m1} \cup B_{f1} = B_T, \ B_{m2} \cup B_{f2} = B_T, \ B_{m3} \cup B_{f3} = B_T \quad (4)$$

The femtocells are to be of plug and play manner. Thus we propose an effective mechanism based on local signalling between femtocells to allocate frequency bands among them which ensures the maximum reuse of available spectrum. A newly installed femtocell must choose its frequency band by sensing its surrounding environment based on the received signals from macrocells and number of interfering femtocells.

**Algorithm 1** Mechanism for selection of frequency band for a newly installed femtocell

1: Detect the received signal $S_1, S_2, S_3$ by the femtocell from the 3 macrocell in the cluster
2: **if** $S_1 > S_2$ and $S_1 > S_3$, **then**
3: the femtocell is within macrocell 1 coverage area;
4: **while** the detected frequency band of MUE = $B_{m1}$ **do**
5: the total frequency band allocation for all femtocells in the macrocell = $B_{f1}$;
6: **if** the signal received by the newly installed femtocell $S_2$ or $S_3$ is greater than $S_{th}$ **then**
7: the femtocell is allocated frequency band = $B'_{f1,celledgefemtocell}$
8: **else if**
9: the femtocell is allocated frequency band = $B'_{f1,innerfemtocell}$ ;
10: **end if**
11: **end while**
12: **while** frequency allocation for femtocell is $B'_{f1,celledgefemtocell}$ **do**
13: search for number of interfering femtocell;
14: **if** the newly installed femtocell does not detect any interfering femtocell **then**
15: $B_{fn} = B_Y + B_Z$;
16: **end if**
17: **if** the newly installed femtocell detects one interfering femtocell **then**
18: $B_{fa} = B_Y + B_X/2$, $B_{fb} = B_Z + B_X/2$;
19: **end if**
20: **if** the newly installed femtocell detects two interfering femtocell **then**
21: $B_{fa} = B_X$, $B_{fb} = B_Y$, $B_{fc} = B_Z$;
22: **end if**
23: **end while**
24: **while** frequency allocation for femtocell is $B'_{f1,innerfemtocell}$ **do**
25: search for number of interfering femtocell;
26: **if** the newly installed femtocell does not detect any interfering femtocell **then**
27: $B_f = B_{m2} + B_{m3}$;
28: **else if** the newly installed femtocell detect one interfering femtocell **then**
29: $B_{fa} = B_{m2}$, $B_{fb} = B_{m3}$;
30: **else if** the newly installed femtocell detect interfering femtocell **then**
31: $B_{fa} = B_{m4}$, $B_{fb} = B_{m5}$, $B_{fc} = B_{m6}$;
32: **end if**
33: **end while**

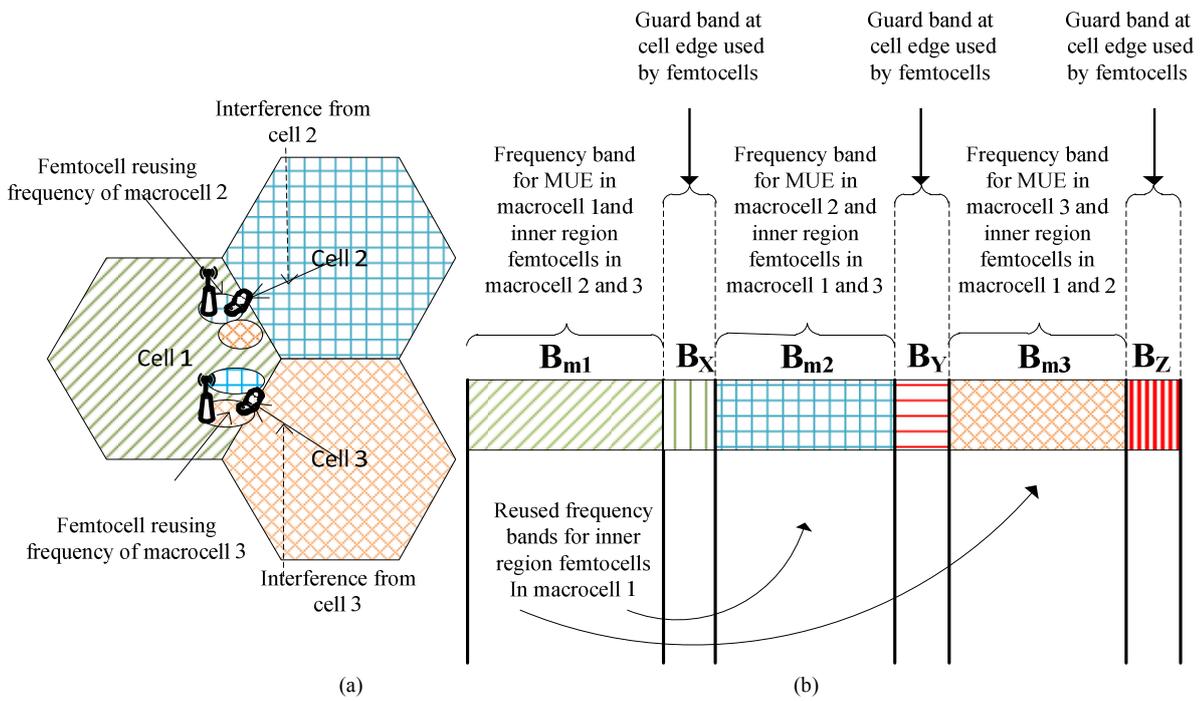

**Figure 1** Frequency allocation with DFR in a cluster of three macrocell. (a) Interference scenario at cell-edge. (b) Proposed frequency planning for macrocell 1.

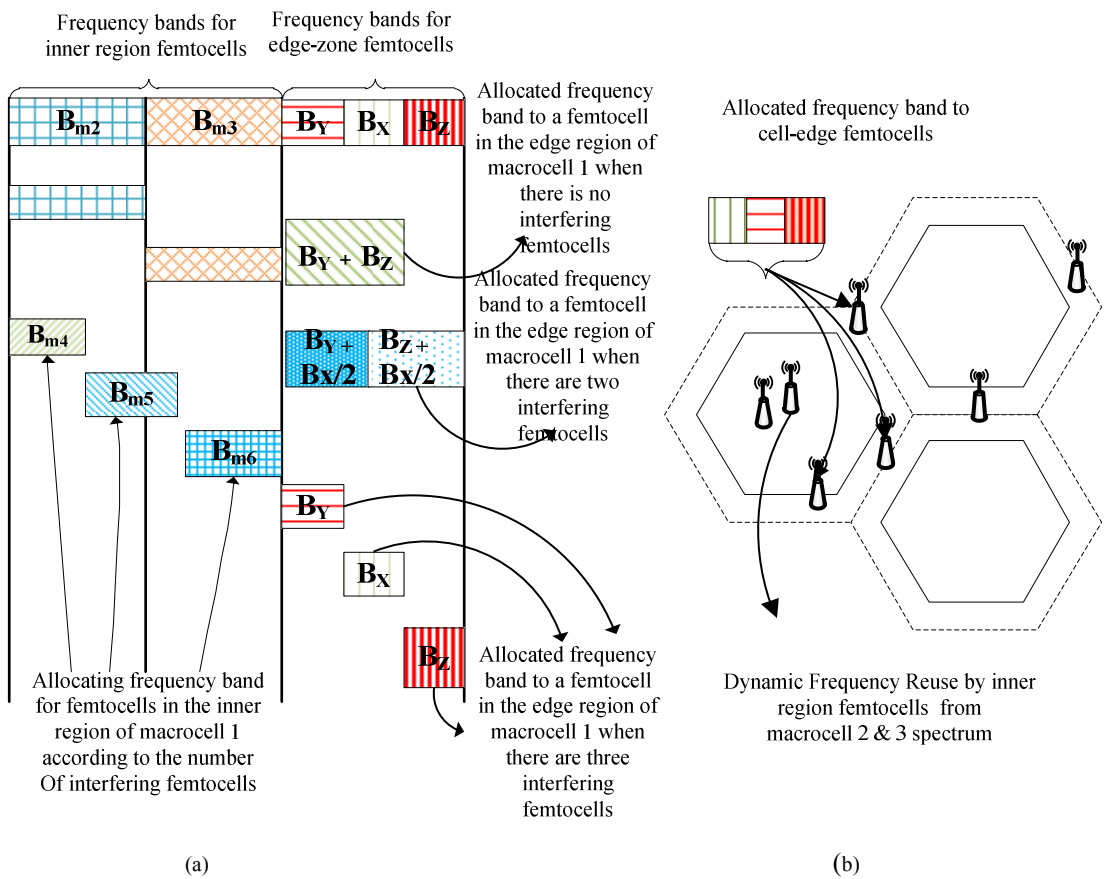

**Figure 2** Proposed DFR scheme. (a) Frequency allocation between femtocells. (b) Allocation of dedicated frequency bands for cell-edge femtocells.

## III. PERFORMANCE ANALYSIS

The performance analysis of our proposed scheme are done based on SINR level, and outage probability. As the validity of interference management schemes are only optimal if they can provide higher SINR level as well as lower outage probability. Okumura-Hata model for cellular path loss calculation have been used in literature [7], [8]. The propagation model for macrocell users can be expressed as:

$$L = 69.55 + 26.16 \log_{10} f_{c,m} - 13.82 \log_{10}(h_b) - a(h_m) \\ + [44.9 - 6.55 \log_{10} h_b] \log_{10} d + L_{sh} \text{ [dB]} \quad (5)$$

$$a(h_m) = 1.1[\log_{10} f_{c,m} - 0.7]h_m - (1.56 \log_{10} f_{c,m} - .8) \quad (6)$$

where $L$ is the path loss, $f_{c,m}$ is the center frequency in MHz of the macrocell, $h_b$ is the height of the macrocellular BS in mater, $h_m$ is the height of the MS in meter, $d$ is the distance between the macrocellular BS and the MS in kilometre, $L_{sh}$ is the shadowing standard deviation.

We consider the FUE to be located in the same indoor environment as the femtocell, then the propagation model can be expressed as:

$$L_{femto} = 20 \log_{10} f_{c,f} + N \log_{10} d_1 - 28 \text{ [dB]} \quad (7)$$

where, $f_{c,f}$ is the center frequency in MHz of the femtocell, $d_1$ is the distance between the femto access point (FAP) and the MS in meter.

The received signal by a MS is given by:

$$P_0 = P * 10^{\frac{-L_{pathloss}}{10}} \quad (8)$$

We consider the spectrums of the transmitted signals to be spread thus the interference can be approximated as AWGN. According to the Shannon Capacity formula:

$$C = W \log_2(1 + SINR) \text{ [bits/s]} \quad (9)$$

Now the received SINR level by a user (of both layer) in a macrocell/femtocell integrated network can be expressed as:

$$SINR = \frac{P_0}{X \sum_{i=0}^{M-1} I_{m(i)} + Y \sum_{j=1}^{K} I_{n(j)} + N} \quad (10)$$

where, $P_0$ is the power of the signals from the associated macrocellular BS or FPA, $I_{m(i)}$ is the power of the interference signal from the $i$-th interfering macrocell from among the $M$ interfering macrocells, and $I_{n(j)}$ is the received interference signal from the $j$-th femtocell from among the $K$ neighbouring femtocells. $N$ is the noise generated by various sources. Here $X$ and $Y$ represents the probability of interference from macrocells and femtocells respectively.

The outage probability of a user is defined as:

$$P_{out} = P_r(SINR < \zeta) \quad (11)$$

where, $\zeta$ is the threshold value of $SINR$ below which there is no acceptable reception.

Considering all the interfering neighbour macrocells and femtocells, the outage probability can be expressed as:

$$P_{out} = 1 - e^{\left(\frac{-\zeta}{SINR}\right)} \quad (12)$$

Sum rate achieved by users within the cell is given by:

$$C_{UE} = \Delta B \log_2(1 + SINR) \quad (13)$$

where, $\Delta B$ is the sub-carrier spacing.

Thus the average system sum rate is given by:

$$C_{avg} = \frac{\sum_{i \in M_i} \sum_{w_m \in W} X_{i \in M_i}^{w_m} C_{i \in M_i}^{w_m}}{\sum MUE} \\ + \frac{\sum_{j \in F_{inner}} \sum_{w_{fin} \in W} X_{j \in F_{inner}}^{w_{fin}} C_{j \in F_{inner}}^{w_{fin}}}{\sum FUE_{inner}} \\ + \frac{\sum_{k \in F_{edge}} \sum_{w_{fedge} \in W_{1,2,3}} X_{k \in F_{edge}}^{w_{fedge} \in W_{1,2,3}} C_{k \in F_{edge}}^{w_{fedge} \in W_{1,2,3}}}{\sum FUE_{edge}} \quad (14)$$

## IV. PERFORMANCE EVALUATION

In this section, we analyse the performance of our proposed model. First for non-dense femtocell deployment then for dense femtocell deployment. Comparisons are done between our proposed scheme and the co-channel deployment, and the hybrid channel allocation schemes described in [5], [6]. Table II shows the parameter values used in our analysis.

Our performance analysis revolves around the number of femtocells deployed in close range. Due to low transmit power and large path losses only those femtocells within 60 meter range of reference femtocell are considered as potential source of interference. First we evaluate the performance of a dense scenario according to our proposed model, and hybrid model.

TABLE II Summary of the simulation parameters

| Parameter | Value |
|---|---|
| Macrocell radius | 1 km |
| Femtocell radius | 10 m |
| Carrier frequency | 900 MHz |
| Macrocell base station transmit power | 1.5 KW |
| Femto base station transmit power | 10 mW |
| Macro base station height | 50 m |
| Femto base station height | 2 m |
| SINR threshold $\zeta$ | 7 (dB) |

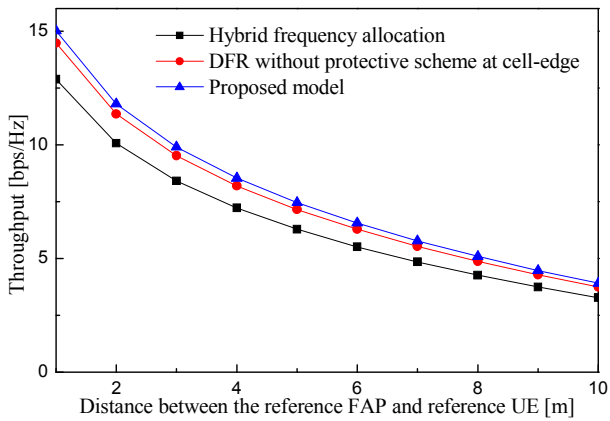

**Figure 3** Throughput comparison for dense femtocell deployment scenarios.

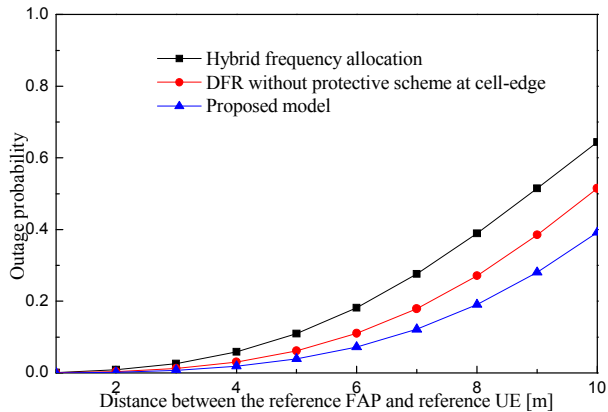

**Figure 4** Outage probability comparison for dense femtocell deployment scenarios.

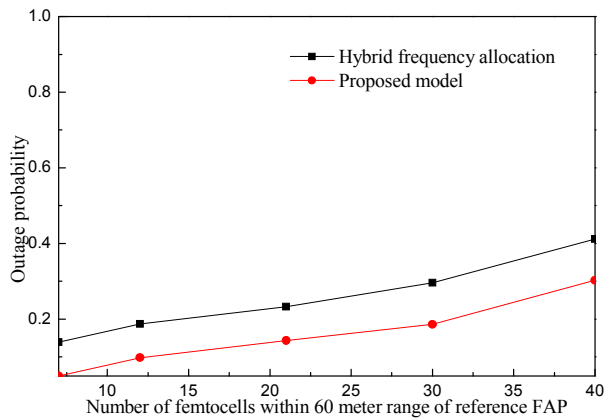

**Figure 5** Outage probability comparison for highly dense femtocell deployment.

We assume a femtocell 900 meter away from the reference macro BS. So the FUE is at the cell-edge region. We also assume that 15 femtocells are being deployed within 60 meter of reference FAP. Fig. 3 shows the throughput comparison. We can see that by using dynamic frequency reuse the capacity of the network has improved. Resultant plots also indicates the necessity of using protective scheme proposed in this paper. Creation of guard region for the cell edge femto users further improves the throughput. Fig. 4 shows the outage probability analysis for dense scenarios. Which confirms the lower outage probability for our proposed scheme.

Next we have considered dense deployment of femtocells (up to 40 femtocells being deployed within 60 meter of reference femtocells). Fig. 5 shows the outage probability comparison. From the above analysis we can summarize that for highly dense deployment of femtocells our model provides higher throughput, and lower outage probability. At the same time ensures higher frequency reuse.

## V. CONCLUSIONS

Wireless networks integrated with small cells might have the answers of capacity enhancement in future generation networks. Proper interference co-ordination can fulfil all the promises made by integrated macro-femto networks. Our work provides an effective way to mitigate interference as well as ensure maximum reuse of macrocell spectrum among femtocells. The detailed frequency planning makes our model capable of ensuring highly dense femtocell deployment without compromising service quality at user level. Our future research includes load balancing, self- healing schemes for further capacity improvement in small cell networks.